# Spin memristive systems

Yuriy V. Pershin[1] & Massimiliano Di Ventra[1]

*[1]Department of Physics, University of California San Diego, La Jolla, California 92093-0319, USA*

**Recently, in addition to the well-known resistor, capacitor and inductor, a fourth passive circuit element, named memristor, has been identified[1] following theoretical predictions.[2,3] The model example used in such case consisted in a nanoscale system with coupled ionic and electronic transport. Here, we discuss a system whose memristive behaviour is based entirely on the electron spin degree of freedom which allows for a more convenient control than the ionic transport in nanostructures. An analysis of time-dependent spin transport at a semiconductor/ferromagnet junction provides a direct evidence of memristive behaviour. Our scheme is fundamentally different from previously discussed schemes of memristive devices and broadens the possible range of applications of semiconductor spintronics.**

In 1971 Leon Chua, analyzing mathematical relations between pairs of fundamental circuit variables, proposed a fourth two-terminal circuit element characterized by a relationship between the charge and the flux[2]. He called that element a memristor (or memory resistor). Five years later, Chua and Kang introduced a more general class of two-terminal devices: memristive systems[3]. If $w$ denotes a set of $n$ state variables describing the internal state of the system, an $n$th-order current-controlled memristive system is described by the equations

$$V = R(w,i,t)i \tag{1}$$

$$\dot{w} = f(w,i,t) \tag{2}$$

where $V$ and $i$ denote the voltage and current, and $R$ is a generalized resistance. The equation for a memristor is a particular case of Eqs. (1) and (2), when $R$ depends only on charge, namely

$$V = R(q(t))i \ . \tag{3}$$

Although several experimental systems were suggested to behave as memristive systems[3], the real interest in these devices resurfaced only recently when a group of scientists from Hewlett Packard have identified a specific experimental realization of memristor[1]. In their scheme, a memory effect is achieved in solid-state thin film two-terminal devices. In that case, the memristive behaviour is based on the coupling between transport of atomic degrees of freedom (e.g., oxygen vacancies acting as mobile dopants defining the internal state of the device) and of electrons. The electron current flowing through such a device dynamically changes its internal state which, in turn, influences the electron transport in a non-linear way.

Ionic transport, however, relies on microscopic features of the system which are generally difficult to control, especially at the nanoscale. The use of a fundamentally different degree of freedom which allows for the realization of memristive behaviour would be thus desirable.

In this work we demonstrate that such degree of freedom is provided by the electron spin and show that memristive behaviour is common for the broad class of semiconductor spintronic devices. This class involves systems whose transport properties depend on the level of electron spin polarization in a semiconductor which is influenced by an external control parameter (such as an applied voltage). The adjustment of electron spin polarization to a variation of the control parameter takes some time (this process involves diffusion and/or relaxation of electron spin polarization) resulting in a memory effect, accompanied by all other requirements for memristive behaviour, such as absence of energy storage. As an example of semiconductor spintronic system exhibiting memristive behaviour, let us consider a semiconductor/half-metal junction as schematically shown in Fig. 1(a). We consider a junction with half-metals (ferromagnets with 100% spin-polarization at the Fermi level) because these act as perfect spin-filters and, therefore, are more sensitive to the level of electron spin polarization. However, we also expect that the prediction of memristive behaviour is valid for junctions with ferromagnets having less than 100% spin polarization as a result of anticipated spin-based peculiarity in the $i-V$ curve of these systems[4]. Understanding the properties of the systems discussed in this Letter is of great interest in the context of actively studied spin-injection/spin-extraction processes at semiconductor/ferromagnet junctions[4-13].

We will be interested mainly in the process of spin extraction, where the electron flow is from the semiconductor into the half-metal, which is especially interesting because of the recently predicted spin-blockade phenomenon in such junctions[4,5]. The physics of the spin-blockade is the following: the half metal accepts electrons of only one, let us say up, spin direction. Spin-down electrons can not enter the half-metal and, therefore, form a cloud near the contact (see the inset in Fig. 2(b)) when a current flows through the system. This cloud increases with increasing current. At a critical current density the density of spin-up electrons near the contact becomes insufficient to provide a further current increase. In other words, transport of spin-up electrons through the contact becomes blocked by the cloud of spin-down electrons near the contact. It was predicted by the present authors[4] that the spin blockade leads to a saturated $i$-$V$ curve as that shown in Fig. 2(b). Here we show that such a system has all the necessary components to exhibit memristive behaviour.

In order to simulate current-voltage characteristics of the junction we assume that the applied voltage mainly drops on the semiconductor part and contact regions. Taking a constant conductivity of the semiconductor region and contact conductivity proportional to spin-up electron density near the contact[4], we can write

$$V = V_s + V_c = \left[ \rho_s L + \rho_c^0 \frac{N_0}{2n_\uparrow(0)} \right] i \qquad (4)$$

where $\rho_s$ and $L$ are the semiconductor resistivity and length, $\rho_c^0$ is the contact resistivity at $V \to 0$, $N_0$ is the electron density in the semiconductor, and $n_\uparrow(0)$ is the density of spin-up electrons near the contact. For simplicity, we assume constant electron density in the semiconductor, i.e., $n_\uparrow + n_\downarrow = N_0$.

The electron spin densities in the semiconductor are described by the spin drift-diffusion model[14]

$$e\frac{\partial n_{\uparrow(\downarrow)}}{\partial t} = \mathrm{div}\vec{i}_{\uparrow(\downarrow)} + \frac{e}{2\tau_{sf}}\left(n_{\downarrow(\uparrow)} - n_{\uparrow(\downarrow)}\right) \quad , \qquad (5)$$

$$\vec{i}_{\uparrow(\downarrow)} = \sigma_{\uparrow(\downarrow)}\vec{E} + eD\nabla n_{\uparrow(\downarrow)} \quad . \qquad (6)$$

Here, $\tau_{sf}$ is the spin relaxation time, $\sigma_{\uparrow(\downarrow)}$ is the conductivity of spin-up (-down) electrons, and $D$ is the diffusion coefficient. According to the present classification scheme for memristive behaviour[3], it is clear that the semiconductor/half-metal junction is a current-controlled memristive system. This can be seen by comparing directly Eqs. (4) and (5) with Eqs. (1) and (2), respectively (we can certainly write Eq. (5) in terms of n discrete variables). In order to observe memristive behaviour, we solved Eqs. (4)-(6) self-consistently at every given time[15] with the following boundary conditions: $i_\uparrow(0) = i$, $i_\downarrow(0) = 0$, $i_\uparrow(L) = i_\downarrow(L) = i/2$ corresponding to the process of spin extraction at the contact.

Fig. 2 shows results of our simulations with a time-dependent applied voltage $V = V_1 + V_2 \sin(2\pi\nu t)$ of high (1GHz) and low (20MHz) frequency $\nu$. Such voltage profile has been selected in order to be always in the spin-extraction regime. The calculated $i-V$ curves (bottom panels in Fig. 2) exhibit a frequency-dependent hysteretic behaviour typical of memristive systems. In particular, we can readily notice the distinctive zero-crossing property of these curves: no current flows through the structure when the voltage drop is zero ($i-V$ curves pass through the $i=0$, $V=0$ point). This property is related to the fact that there is no energy storage in our device as opposed to the energy storage in the usual capacitive or inductive circuit elements.

Another interesting feature of the $i-V$ curves is their frequency behaviour. It follows from Fig. 2 that the hysteresis is significantly suppressed at 20MHz frequency. This is a manifestation of the fact that at low frequencies our system behaves essentially as a non-linear resistor. Physically, at low applied voltage frequencies, the electron spin polarization in the semiconductor has enough time to adjust to any present value of the voltage. Therefore, the current through the system at low frequencies is non-linear (because of spin-blockade) but essentially history-independent (no hysteresis). It is also clear that at very high frequencies ($\upsilon \to \infty$) the electron spin polarization does not have any time for redistribution within the oscillation period. Therefore, the spin-up density near the contact and, correspondingly, the contact resistivity can be considered constant, so that the device operates as a linear resistor. The above described frequency behaviour of the semiconductor/half-metal junction is typical for memristive systems[3].

Although the model represented by Eqs. (4)-(6) takes into account the main physics of the underlying device operation, it is still quite complex for circuit analysis. Here, we thus discuss a simplified (and more transparent) model which can be obtained in the following way. Instead of tracking the whole spin density distribution in the semiconductor, let us focus our attention on the integrated spin density (surface spin density) $N_{\uparrow(\downarrow)}=\int_0^L n_{\uparrow(\downarrow)}dx$ and select this quantity as the only system parameter. Integrating Eq. (5) from 0 to $L$ (i.e., over the whole semiconductor length) we obtain the following equation

$$e\frac{\partial N_\uparrow}{\partial t}=i_\uparrow(L)-i_\uparrow(0)+\frac{e}{2\tau_{sf}}\left(N_\downarrow-N_\uparrow\right)=-\frac{1}{2}i+\frac{e}{\tau_{sf}}\left(\frac{NL}{2}-N_\uparrow\right) \quad . \tag{7}$$

Eq. (7) simply states that the change of the integrated spin-up density is due to the injection/extraction of spin-up electrons through the semiconductor boundaries (the first "current" term in the RHS of Eq. (7)) and spin relaxation processes (the second term in the RHS of Eq. (7)). Although the knowledge of $N_\uparrow$ is not sufficient to exactly obtain $n_\uparrow(0)$, which enters into Eq. (4), we can approximately write

$$n_\uparrow(0)=f(N_\uparrow), \tag{8}$$

where $f$ is a given smooth function. The simplified circuit model given by Eqs. (4), (7), (8) describes again a memristive system. However, at short times, when spin relaxation processes are not important the last term in the RHS of Eq. (7) can be neglected. The resulting set of equations describes then a perfect memristor. Indeed, integrating Eq. (7) we obtain

$$N_\uparrow(t)=N_\uparrow(0)-\frac{1}{2e}\int_0^t i(\tau)d\tau=N_\uparrow(0)-\frac{1}{2e}q(t) \tag{9}$$

and, substituting Eqs. (8) and (9) into Eq. (4), we get

$$V=\left[\rho_s L+\rho_c^0\frac{N_0}{2f\left(N_\uparrow(0)-\frac{1}{2e}q(t)\right)}\right]i \quad . \tag{10}$$

It follows from Eq. (10), which is a particular case of Eq. (3), that the resistivity of the system depends only on the amount of charge $q(t)$ flowing through it.

In order to test the simplified circuit model predictions, we calculate, using Eqs. (4)-(6), the short-time system response to unipolar and bipolar voltage steps excitations. Fig. 3 (b) demonstrates that our structure behaves almost as a perfect memristor at short times. In particular, the red line in Fig. 3 (b) shows that when the total charge flowing through the structure is equal to zero, the value of the system parameter $n_\uparrow(0)$ (and consequently of the device resistivity) becomes very close to its initial value already at $t = 5$ ps. The opposite behaviour is demonstrated by the blue line in Fig. 3 (b): when the total charge flowing in the system is different from zero, the value of the spin-up density at the interface $n_\uparrow(0)$ is quite different from its initial value even at $t = 10$ ps.

To summarize, we have demonstrated that a semiconductor/half-metal junction is in fact a memristive system. The origin of its unusual behaviour in a circuit is completely based on the electronic spin degree of freedom, which is much easier to control than ionic transport at the nanoscale. In our scheme, the principal role is played by electron spin diffusion and relaxation processes which drive the system to equilibrium. Moreover, we would like to note that the spin memristive behaviour is not limited to the discussed device; it should be typical of many semiconductor spintronic devices that involve spin-filters, general semiconductor-ferromagnet junctions, etc. We also expect the memristive behaviour to be more pronounced in structures with low electron density, where the level of electron spin polarization can be significantly varied by external control parameters. We thus believe that our demonstration of memristive effects in spintronics is an important step forward in future practical applications of the newly discovered fourth circuit element.

**References**

1. Strukov, D. B., Snider, G. S., Stewart, D. R. & Williams, R. S. The missing memristor found. *Nature* **453**, 80-83 (2008)

2. Chua, L. O. Memristor - the missing circuit element. *IEEE Trans. Circuit Theory* **18**, 507–519 (1971)

3. Chua, L. O. & Kang, S. M. Memristive devices and systems. *Proc. IEEE* **64**, 209–223 (1976)

4. Pershin, Y. V. & Di Ventra, M. Current-voltage characteristics of semiconductor/ferromagnet junctions in the spin-blockade regime. *Phys. Rev. B* **77**, 073301 (2008)

5. Pershin, Y. V. & Di Ventra, M. Spin blockade at semiconductor/ferromagnet junctions. *Phys. Rev. B* **75**, 193301 (2007)

6. Rashba, E.I. Diffusion theory of spin injection through resistive contacts. *Eur. Phys. J. B* **29**, 513-527 (2002).

7. Albrecht, J. D. & Smith, D. L. Spin-polarized electron transport at ferromagnet/semiconductor Schottky contacts. *Phys. Rev.* B **68**, 035340 (2003);

8. Žutić, I., Fabian, J. & Das Sarma, S. Spin-Polarized Transport in Inhomogeneous Magnetic Semiconductors: Theory of Magnetic/Nonmagnetic p-n Junctions. *Phys. Rev. Lett.* **88**, 066603 (2002).

9. Kawakami, R. K. et al. Ferromagnetic Imprinting of Nuclear Spins in Semiconductors. *Science* **294**, 131-134 (2001).

10. Crooker, S. A. et al. Imaging Spin Transport in Lateral Ferromagnet/Semiconductor Structures. *Science* **309**, 2191-2195 (2005).

11. Shen, M., Saikin, S. & Cheng, M.-C. Monte Carlo modeling of spin injection through a Schottky barrier and spin transport in a semiconductor quantum well. *J. Appl. Phys.* **96**, 4319-4325 (2004).

12. Wang, Y. Y. & Wu, M. W. Schottky-barrier-induced spin relaxation in spin injection. *Phys. Rev. B* **72**, 153301 (2005).

13. Dery, H. & Sham, L. J. Spin Extraction Theory and Its Relevance to Spintronics. *Phys. Rev. Lett.* **98**, 046602 (2007).

14. Yu, Z. G. & Flatté, M. E. Electric-field dependent spin diffusion and spin injection into semiconductors. *Phys. Rev. B* **66**, 201202 (2002)

15. Scharfetter, D. L. & Gummel, H. K. Large signal analysis of a silicon read diode oscillator. *IEEE Trans. Electron Devices* **ED-16**, 64-77 (1969)

**Acknowledgements** This work was partially supported by the National Science Foundation grant NSF-0438018 and Department of Energy grant DE-FG02-05ER46204.

## Figure Captions

**Figure 1 | Semiconductor/half-metal junction. a,** Schematic representation of the circuit made of an interface between a semiconductor and a half-metal. **b,** Typical DC current-voltage characteristics. Inset: spin-up and spin-down densities in the semiconductor region as a function of the distance from the contact. $i_c = eN_0\sqrt{D/2\tau_{sf}}$ is the critical current density[5].

**Figure 2 | Simulations of AC response of the system.** The applied voltages (blue lines) are $V = V_1 + V_2\sin(2\pi\nu t)$ with $V_1 = V_2 = 0.5$ V, $\nu = 10^9$ Hz in **a** and $\nu = 2\cdot10^7$ Hz in **b**. In both cases the current densities (green lines – in units of the critical current density) and spin-up electron densities near the contact (red lines) show saturation typical for spin blockade[5]. It is clearly seen that $i-V$ hysterisis is significantly reduced in the low-frequency case. The calculations were made using the following system parameters: $L = 20\mu\mathrm{m}$, $D = 220\mathrm{cm}^2/\mathrm{s}$, $\mu = 8500\mathrm{cm}^2/(\mathrm{Vs})$, $\tau_{sf} = 2\mathrm{ns}$, $N_0 = 5\cdot10^{15}\mathrm{cm}^{-3}$ and $\rho_c^0/(\rho_s L) = 1$.

**Figure 3 | System dynamics excited by step voltages. a,** Unipolar (blue line) and bipolar (red line) step voltage profiles used in our calculations. The spin-injection process (at negative applied voltage) was modeled using a constant interface resistance model. **Inset:** Total charge flowing through the system as a function of time. The profile of bipolar voltage was selected in such a way that the corresponding total charge $q = 0$ for $t > 2$ ps (see the red line). **b,** Evolution of spin-up electron density near the interface. In the case of bipolar voltage excitation (red line), the final value of spin-up density is close to its initial value (shown by the dotted line). This is a manifestation of a nearly perfect memristor behavior.

Figure 1

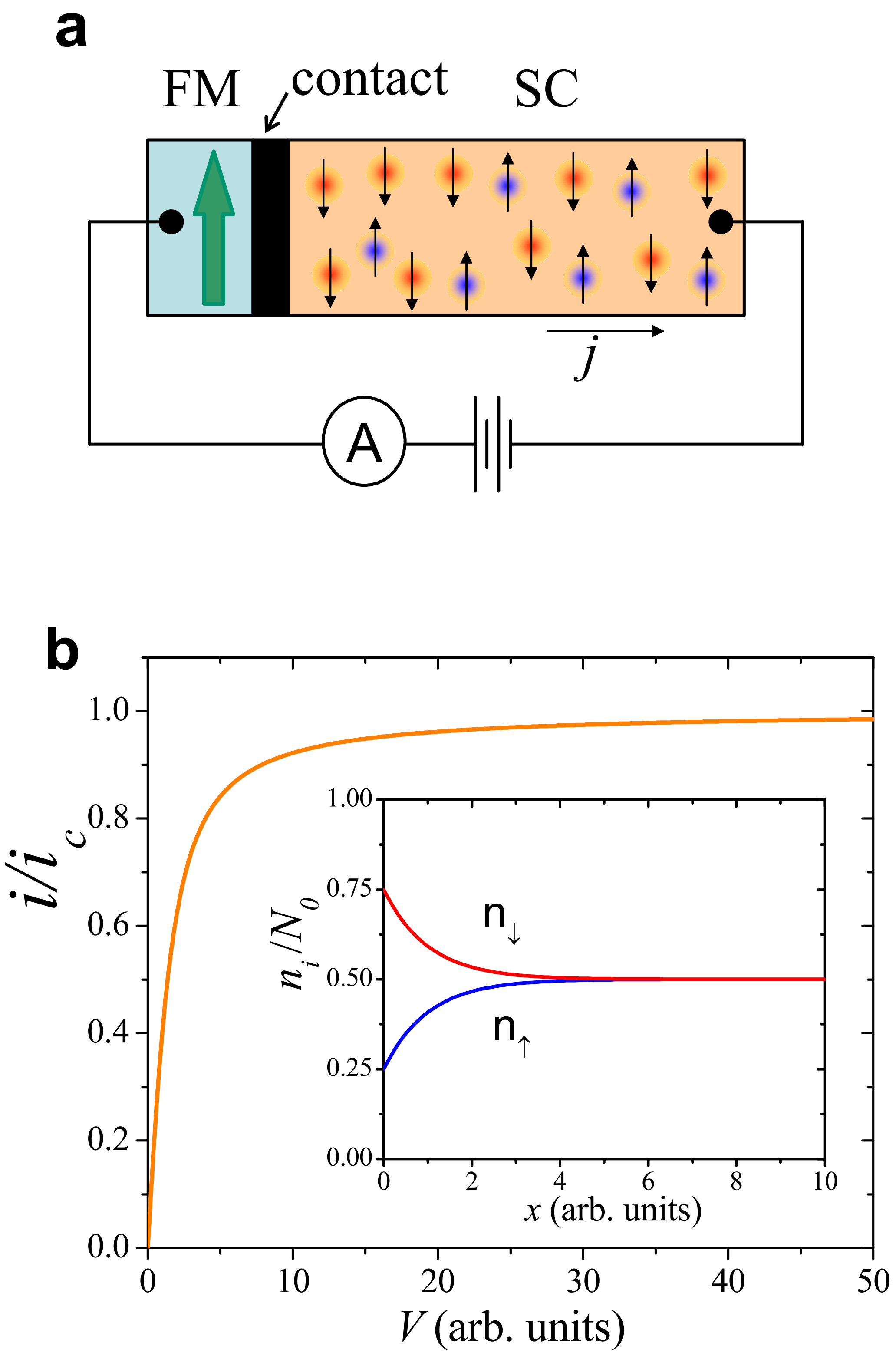

Figure 2

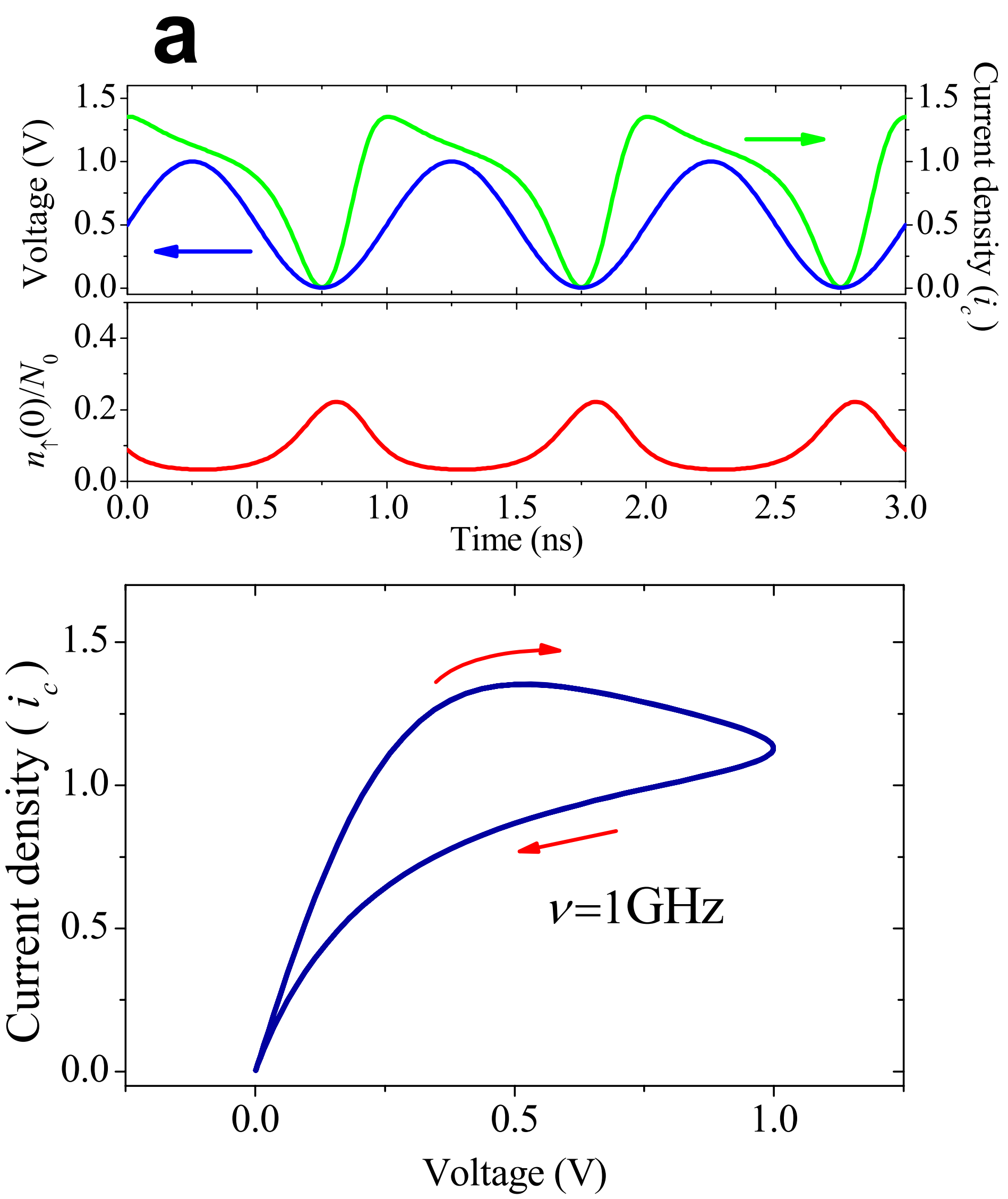

Figure 2

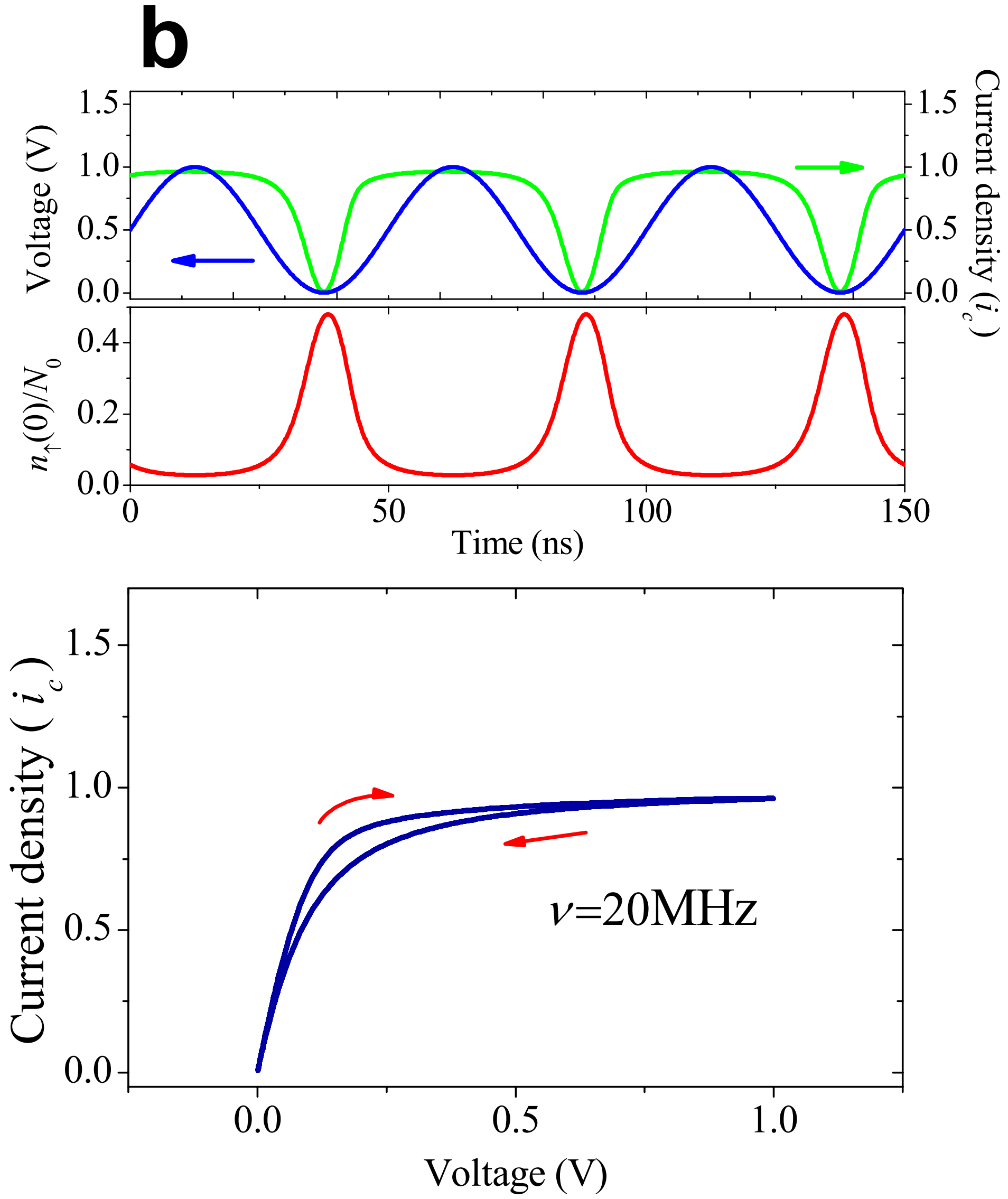

Figure 3

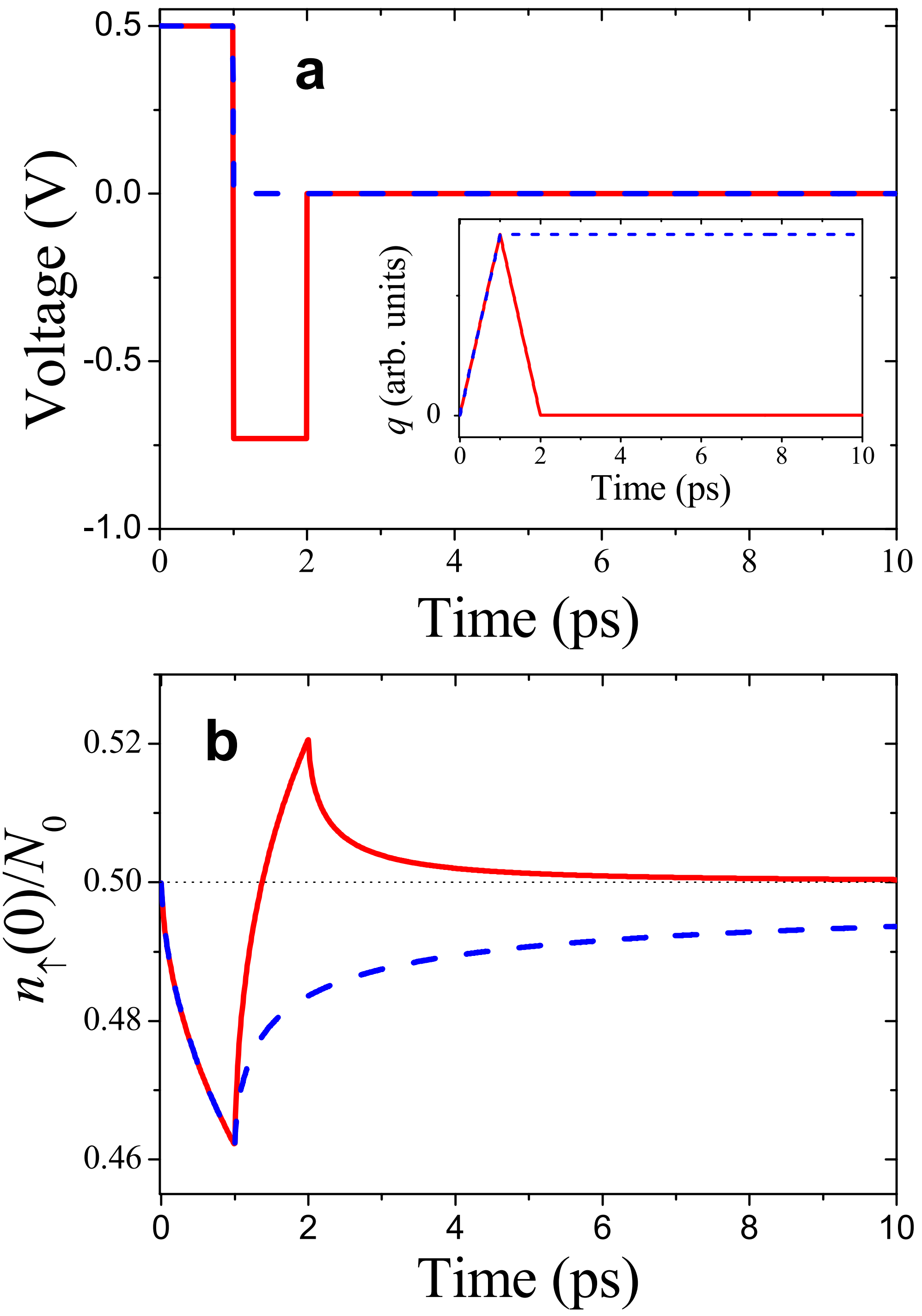